\documentclass[11pt,leqno]{article} %11pt,
\usepackage{amsmath,amstext, amsthm, amssymb}

 \numberwithin{equation}{section}
\newtheorem{thm}{Theorem}[section]
\newtheorem{lem}[thm]{Lemma}

\newcommand{\al}{\alpha}

\newcommand{\W}{\frac{u(qx)-u(y)}{qx-y}}

\newcommand{\be}{\begin{equation}}
\newcommand{\ee}{\end{equation}}
\newcommand{\bea}{\begin{eqnarray}}
\newcommand{\eea}{\end{eqnarray}}

\newcommand{\ba}{\begin{array}}
\newcommand{\ea}{\end{array}}

\newcommand{\bt}{\beta}

\newcommand{\ints}{\int_{0}^{\infty}}
\newcommand{\non}{{\nonumber}}

\newcommand{\dqi}{D_{q^{-1}}}
\newcommand{\dq}{{D_q}}
\newcommand{\sqq}{{\sqrt {q}}}
\begin{document}

\title{Ladder Operators for $q$-orthogonal Polynomials
\thanks{This work started during the authors' visit to the Institute of Mathematical Sciences in Singapore and the support
from the institute is acknowledged}}
\author{Yang Chen\\
         Department of Mathematics\\
          Imperial College\\
          180 Queen's Gate\\
          London SW7 2BZ, UK\\
          ychen@ic.ac.uk\\
          \and Mourad E.H. Ismail
\\ Department of Mathematics \\ University of Central Florida
\\  Orlando, FL 32816\\
USA\\
ismail@math.ucf.edu}
%\date{}

\maketitle

\begin{abstract}
The $q-$ difference analog of the classical ladder operators is derived for those orthogonal
polynomials arising from a class of indeterminate moments problem.
\end{abstract}

% \vfill\eject
%%%%%%%%%%%%%%%%%%%%%%%%%%%%%%%%%%%%%%%%%%%%%%%%%%%%%%%%%%%%%%%%%%%

 \setcounter{equation}{0}
 \setcounter{thm}{0}
\section{Introduction}
This work is a follow up to our work \cite{Che:Ism} where we derived raising and lowering
operators for polynomials orthogonal with respect to absolutely continuous measures $\mu$ under certain
smoothness assumptions of  $\mu'$. This approach goes back to \cite{Bau}, \cite{Bon:Cla}, and \cite{Mha}.
The raising and lowering operators derived in these references are differential operators.
It was later realized that a similar theory exists for polynomials orthogonal with respect
to a measure with masses at the union of at most two geometric progressions, $\{a q^n, bq^n\}$, for some $q \in (0,1)$,
\cite{Ism8}. The corresponding theory for difference operators is in \cite{Ism:Nik:Sim}.
This material is reproduced in \cite{Ismbook}.  The raising and lowering operators involve two
functions $A_n(x)$ and $B_n(x)$ which satisfy certain recurrence relations. In the case of differential
 operators we have demonstrated that the knowledge of $A_n(x)$
and $B_n(x)$ determines the polynomials uniquely in the cases of Hermite, Laguerre,
and Jacobi polynomials, see \cite{Che:Ism2}. This is done through recovering the properties of the polynomials
including the three term recurrence relation which generates the polynomials. This work shows that the
corresponding functions determine the polynomials in the cases of Stieltjes-Wigert and $q$-Laguerre polynomials.

The orthogonal polynomials which arise from indeterminate moment problems have
discrete and absolutely
continuous orthogonality measures \cite{Akh}.
In many instances it is more convenient to work with
absolutely continuous measures
\cite[Chapter 21]{Ismbook}.

In this work we derive raising and lowering operators for
polynomials orthogonal with respect to absolutely continuous
measures. We shall assume that $\{P_n(x)\}$ are monic orthogonal
polynomials, so that
\bea
\label{eqdefofu}
\label{eqorth.rel.} \int_{0}^{\infty}
w(x)P_{m}(x)P_{n}(x)dx = \zeta_{n}\delta_{m,n}.
 \eea
 A weight function
$w$ leads to a potential $u$ defined by
\bea
\label{eqdefu}
 u(x)=
-\frac{\dqi w(x)}{w(x)},
\eea
where $\dq$ is the $q$-difference
operator
\bea
\label{eqdefDq}
(\dq f)(x) = \frac{f(x)-f(qx)}{x-qx}.
\eea
Every monic sequence of orthogonal polynomials satisfies a
three term recurrence relation of the form
\bea
 \label{eq3trr}
(x-\al_n)P_n(x) =
P_{n+1}(x) + \beta_n P_{n-1}(x).
\eea
We also write the monic polynomials $P_n(x)$ as follows:
\bea
P_n(x)=x^{n}+\textsf{p}_1(n)x^{n-1}+...\non
\eea
and it follows immediately from the three term recurrence relations (1.4) that
\bea
\label{eqp1intermsofalpha}
\al_n=\textsf{p}_1(n)-\textsf{p}_1(n+1).
\eea
A main result
of this work is the following theorem.
\begin{thm}
Let
\begin{gather}
\label{eqdefnAn}
A_n(x) := \frac{1}{\zeta_n}\int_0^\infty \W P_n(y)P_n(y/q)\, w(y) dy, \\
\label{eqdefnBn} \quad B_n(x) := \frac{1}{\zeta_{n-1}}\int_0^\infty \W
P_n(y)P_{n-1}(y/q)\, w(y) dy.
\end{gather}
Then we have the lowering relation
\bea
\label{eqlowering}
\dq P_n(x) = \bt_n A_n(x)P_{n-1}(x)  -B_n(x)P_n(x).
\eea
\end{thm}
Theorem 1.1 will be proved in \S2 along with the difference equations satisfied by
$A_n(x)$ and $B_n(x);$
\begin{gather}
\label{eqqS_1}
B_{n+1}(x)+B_n(x)=(x-\al_n)A_n(x)+x(q-1)\sum_{j=0}^{n}
A_j(x)-u(qx),  \\
1+(x-\al_n)B_{n+1}(x)-(qx-\al_n)B_n(x)
=\bt_{n+1}A_{n+1}(x)-\bt_{n}A_{n-1}(x).
\label{eqqS_2}
\end{gather}
The identities \eqref{eqqS_1}--\eqref{eqqS_2} will be referred to as the supplementary conditions.

Theorem 1.1 is the $q$-analogue of
$$P_n^\prime(x) = \bt_n A_n(x)P_{n-1}(x)  -B_n(x)P_n(x)$$
of \cite{Che:Ism}. See also \cite{Che:Ism2} for a derivation of the
supplementary conditions for the $q=1$ case.

Let
\bea
L_{1,n} := \dq + B_n(x),
\label{eqdefnL1n}
\eea
be the lowering operator. Thus \eqref{eqlowering} is
\bea
\label{eqlowering2}
L_{1,n} P_n(x) = \bt_nA_n(x) P_{n-1}(x).
\eea
The raising operator can be found as follows: First replace $\bt_nP_{n-1}(x)$ in
(1.8) by $(x-\al_n)P_{n}(x)-P_{n+1}(x),$ using (1.4), and
then replace
$$
-(x-\al_n)A_n(x)+B_n(x)
$$
by
$$
x(q-1)\sum_{j=0}^{n}A_j(x)-u(qx)-B_{n+1}(x),
$$
using (1.9). An easy computation shows that,
$$
D_qP_{n}(x)=\left(B_{n+1}(x)+u(qx)-x(q-1)\sum_{j=0}^{n}A_j(x)\right)P_{n}(x)
-A_{n}(x)P_{n+1}(x).
$$
With the replacement of $n$ by $n-1$ in above equation, the raising operator is,
\bea
L_{2,n}:=D_q+x(q-1)\sum_{j=0}^{n-1}A_j(x)-u(qx)-B_{n}(x),\non
\eea
and
\bea
L_{2,n}P_{n-1}(x)=-A_{n-1}(x)P_{n}(x).\non
\eea
It is useful to recall the  following analogue of the product rule
\bea
\label{eqqproductrule}
\dq(f(x)g(x))=(\dq f(x))g(x)+f(xq)\dq g(x).
\eea
The following lemma, whose proof is a calculus exercise, will be used in the proofs of our main results.
\begin{lem}
If the integrals
$$
\int_0^\infty f(x) g(x) \frac{dx}{x}, \quad \int_0^\infty f(x) g(qx) \frac{dx}{x},
$$
exist then the following $q$-analogue of integration by parts holds
\bea
\label{eqqintegbyparts}
\int_{0}^{\infty}f(x)\dq g(x)dx=-\frac{1}{q}\int_{0}^{\infty}
g(x)\dqi f(x)dx.
\eea
\end{lem}

An immediate consequence of Lemma 1.2 and \eqref{eqorth.rel.}  is
\bea
\label{eqintegraluw}
 \int_{0}^{\infty}u(y)P_n(y)P_n(y/q)w(y)dy =0.
\eea
We also have
\bea
\label{eqintpnpn+1}
 \int_{0}^{\infty}u(y)P_{n+1}(y)P_n(y/q)w(y)dy = \frac{1-q^{n+1}}{1-q}q\zeta_n,
\eea
 which follows from  \eqref{eqqproductrule}, \eqref{eqdefu},  \eqref{eqqintegbyparts},
 and the fact  that
\bea
\dq x^n=\frac{1-q^n}{1-q}x^{n-1}.
\non
\eea
 \section{Proofs}
 We shall need the fact \cite{Sze}, \cite{Ismbook}
 \bea
 \label{eqbeta-zeta}
 \zeta_n =\zeta_0 \bt_1\bt_2 \dots \bt_n,
 \eea
and  the Christoffel--Darboux identity \cite{Sze}, \cite{Ismbook}
 \bea
 \label{eqCD}
 \sum_{k=0}^{n-1} \frac{P_k(x)P_k(y)}{\zeta_k} = \frac{P_n(x) P_{n-1}(y) - P_n(y) P_{n-1}(x)}
 {\zeta_{n-1}(x-y)}.
 \eea

 \begin{proof}[Proof of Theorem 1.1]
 Let $\dq P_n(x)=\sum_{k=0}^{n-1}c_{n,k}P_k(x)$. Then
 $$
\zeta_kc_{n,k} = \int_{0}^{\infty}P_k(y)w(y)\dq P_n(y)dy.
$$
Applying Lemma 1.2, \eqref{eqqproductrule}, we see that
\bea
q\zeta_kc_{n,k} &=&-\int_{0}^{\infty}P_n(y)
\left[(\dqi P_k(y))w(y)+P_k(y/q)\dqi w(y)\right]dy\non\\
&=& \int_{0}^{\infty}P_n(y)P_k(y/q)
\left[-\frac{\dqi w(y)}{w(y)}\right]w(y)dy \non
\eea
where the orthogonality property was used in the last step. The definition of $u$ \eqref{eqdefu} yields
\bea
q\zeta_kc_{n,k}&=&\int_{0}^{\infty}P_n(y)P_k(y/q)u(y)w(y)dy\non\\
&=&-\int_{0}^{\infty}P_n(y)P_k(y/q)(u(qx)-u(y))w(y)dy\non,
\eea
where we again used the orthogonality property in the last step. Therefore
by the Christoffel--Darboux identity \eqref{eqCD}
\bea
\begin{gathered}
\dq P_n(x)\non\\
= -\frac{1}{\zeta_{n-1}}
\int_{0}^{\infty}P_n(y)\frac{u(qx)-u(y)}{qx-y}
\left[P_n(x)P_{n-1}(y/q)-P_n(y/q)P_{n-1}(x)\right]w(y)dy\non
\end{gathered}
\eea
and \eqref{eqbeta-zeta}, the theorem follows.
\end{proof}

\begin{proof}[Proof of \eqref{eqqS_1}] It is clear that
\bea
 \begin{gathered}
  B_{n+1}(x)+B_n(x)\\
= \int_{0}^{\infty}\frac{u(qx)-u(y)}{\zeta_n(qx-y)}
\left[P_{n+1}(y)P_n(y/q)+\bt_nP_n(y)P_{n-1}(y/q)\right]
w(y)dy \\
= I_1 + I_2,  \nonumber
\end{gathered}
\eea where
\bea I_1&:=&\frac{1}{\zeta_n}\int_{0}^{\infty}
\W \left(y/q-\al_n)P_n(y\right)P_n(y/q)w(y)dy\non\\
I_2&:=&\frac{1}{\zeta_n}\int_{0}^{\infty}\W \left[P_{n+1}(y)P_n(y/q)
-P_n(y)P_{n+1}(y/q)\right]w(y)dy, \non
\eea
after
$\bt_nP_{n-1}(y/q)$ is replaced by $(y/q-\al_n)P_n(y/q) -
P_{n+1}(y/q)$. It is easy to see that $I_1$ is given by
\bea
I_1&=&(x-\al_n)A_n(x)-\frac{1}{\zeta_nq}\int_{0}^{\infty}
(u(qx)-u(y))P_n(y)P_n(y/q)w(y)dy\non\\
&=&(x-\al_n)A_n(x)-q^{-n-1}u(qx),  \non \eea
where
\eqref{eqintegraluw} and the fact that
\bea
\label{eqP(y/q)}
P_j(y/q)=q^{-j}P_j(y)+ \textup{lower degree terms}
 \eea
were used. To
evaluate $I_2$ first note that \eqref{eqP(y/q)} implies
\bea
\int_{0}^{\infty} P_j(y)P_j(y/q) w(y)dy= \zeta_jq^{-j}.
\eea
Next we
apply the Christoffel--Darboux formula to
$$P_{n+1}(y)P_n(y/q)-P_n(y)P_{n+1}(y/q), $$
and replace $y -y/q$  by $(y-qx+qx)(1-1/q)$. Therefore we see that
\bea
\begin{gathered}
I_2=x(q-1)\sum_{j=0}^{n}A_j(x) +\frac{1-q}{q} \int_{0}^{\infty}\left[u(qx)-u(y)\right]
\sum_{j=0}^{n}\frac{P_j(y)P_j(y/q)}{\zeta_j}w(y)dy \\
= x(q-1) \sum_{j=0}^{n}A_j(x) + \frac{1-q}{q} u(qx)\sum_{j=0}^{n}
q^{-j} + \frac{1-q}{q}
\int_{0}^{\infty}\sum_{j=0}^{n}\frac{P_j(y)P_{j}(y/q)}{\zeta_j}\dqi
w(y)\, dy.
\end{gathered}
\non
\eea
Thus
$$
I_2=x(q-1)\sum_{j=0}^{n}A_j(x) + \frac{1-q}{q} u(qx)\frac{1-q^{-n-1}}{1-q^{-1}}.
$$
Simplifying $I_1+I_2$ we establish \eqref{eqqS_1}.
\end{proof}
\begin{proof}[Proof of \eqref{eqqS_2}]
From the definition of $B_n(x)$ we see that
\bea
\begin{gathered}
(x-\al_n)B_{n+1}(x)-(qx-\al_n)B_n(x)\\
=\ints w(y)\frac{u(qx)-u(y)}{qx-y}\\
\times
\left[\left(\frac{x-\al_n}{\zeta_n}\right)P_{n+1}(y)P_n(y/q)
-\left(\frac{qx-\al_n}{\zeta_{n-1}}\right)P_n(y)P_{n-1}(y/q)\right]dy\\
=\ints w(y)\left[u(qx)-u(y)\right]\left
[\frac{1}{\zeta_n}\left(\frac{1}{q}+\frac{y/q-\al_n}{qx-y}\right)P_{n+1}(y)P_{n}(y/q)\right. \\
-\left. \frac{1}{\zeta_{n-1}}\left(1+\frac{y-\al_n}{qx-y}\right)P_n(y)P_{n-1}(y/q)]\right]\,
dy \\
=-\frac{1}{\zeta_nq}\ints w(y)u(y)P_{n+1}(y)P_{n}(y/q)dy \\
+\frac{1}{\zeta_n}\ints
w(y)\frac{u(qx)-u(y)}{qx-y}(y/q-\al_n)P_n(y/q)P_{n+1}(y)dy
\\
+\frac{1}{\zeta_{n-1}}\ints w(y)u(y)P_n(y)P_{n-1}(y/q)dy\\
-\frac{1}{\zeta_{n-1}}\ints w(y)\frac{u(qx)-u(y)}{qx-y}P_n(y)P_{n-1}(y/q)(y-\al_n)dy\\
=-\frac{1}{\zeta_nq}\ints w(y)u(y)P_{n+1}(y)P_{n}(y/q)dy\\
+\frac{1}{\zeta_n}\ints
w(y)\frac{u(qx)-u(y)}{qx-y}\left[P_{n+1}(y/q)+\bt_nP_{n-1}(y/q)\right]
P_{n+1}(y)dy\\
+\frac{1}{\zeta_{n-1}}\ints w(y)u(y)P_n(y)P_{n-1}(y/q)dy\\
-\frac{1}{\zeta_{n-1}}\ints
w(y)\frac{u(qx)-u(y)}{qx-y}\left(P_{n+1}(y)+\bt_nP_{n-1}(y)\right)
P_{n-1}(y/q)dy. \non
\end{gathered}
\eea
The result follows after some simplifications using \eqref{eqintpnpn+1}.
\end{proof}

\setcounter{equation}{0}
\section{Stieltjest-Wigert polynomials}

The computations in this and the next section will show that
 $A_n(x)$ and $B_n(x)$ are rational in $x$
and consequently (1.9) and (1.10), the supplementary conditions which are identities
valid for all $x$ would be particularly useful for the recovery of the recurrence
coefficients. As we shall see, systems of apparently non-linear difference equations
generated by equating the residues on both sides of (1.9) and (1.10), can be solved
explicitly which ultimately determine the recurrence coefficients.

This is example of an indeterminate moment problem associated with the
log-normal density. See \cite{Chr03} for a discussion of the associated moment
problem. We take the weight to be
$$
w(x)=c\exp[(\ln x)^2/(2\ln q)],\;\;0<x<\infty,\;\;0<q<1,
$$
where $c$ is a positive constant which will not appear in subsequent computations.

 An easy calculation shows that
\bea
u(x)&=&\frac{q}{1-q}\left(\frac{1}{x}-\frac{{\sqrt q}}{x^2}\right),
\non\\
A_n(x)&=&\frac{R_n}{x^2},\non\\
{\rm where\;\;}R_n&:=&\frac{1}{\zeta_n(1-q)\sqq}\int_{0}^{\infty}
P_n(y)P_n(y/q)w(y)\frac{dy}{y},\non\\
B_n(x)&=&\frac{r_n}{x^2}-\frac{1-q^n}{1-q}\frac{1}{x}\non\\
{\rm where\;\;}r_n&:=&\frac{1}{\zeta_{n-1}\sqq(1-q)}\int_{0}^{\infty}
P_n(y)P_{n-1}(y/q)w(y)\frac{dy}{y}.\non
\eea
From the supplementary conditions, \eqref{eqqS_1} and \eqref{eqqS_2}
\bea
\frac{q^{n+1}+q^n-2}{1-q}&=&R_n+(q-1)S_n-\frac{1}{1-q}\\
r_{n+1}+r_n&=&-\al_n R_n+\frac{1}{\sqq(1-q)}\\
0&=&\al_n\frac{1-q^{n+1}}{1-q}
-\al_n\frac{1-q^n}{1-q}+r_{n+1}-qr_n\\
\al_n(r_n-r_{n+1})&=&\bt_{n+1}R_{n+1}-\bt_nR_{n-1},
\eea
where $S_n:=\sum_{j=0}^{n}R_j,$ and
\bea
R_0
=\frac{1}{\sqq(1-q)}\frac{\int_{0}^{\infty}w(y)/y\;dy}{\int_{0}^{\infty}w(y)dy}
=\frac{1}{1-q}.
\eea
A difference equation satisfied by $R_n$ is found by subtracting (3.1) at $"n-1"$ from
the same at $"n"$;
\bea
qR_n-R_{n-1}=-(1+q)q^{n-1},
\eea
and since the "integrating factor" is $q^{-n},$ the unique solution is
\bea
R_n=\frac{q^n}{1-q}.
\eea
Note that (3.3) simplifies to
\bea
-\al_n q^n=r_{n+1}-qr_n.
\eea
Multiplying (3.2) by $1-q$, together with $R_n(1-q)=q^n$ and (3.8) one
finds
\bea
r_n-qr_{n+1}=\frac{1}{\sqq},
\eea
and since the "integrating factor" for this is $q^n,$
the unique solution subject to $r_0=0,$ is
\bea
r_n=\frac{1-q^{-n}}{(1-q)\sqq},
\eea
which with (3.8) immediately gives,
\bea
\al_n=\frac{q^{-n}}{\sqq}\left(q^{-n-1}+q^{-n}-1\right).
\eea
Multiply (3.4) by $R_n$ and replace $\al_nR_n$ with (3.2), we find the resulting
first order difference equation
\bea
r_{n+1}^2-\frac{r_{n+1}}{{\sqrt{q}}(1-q)}-\left(r_{n}^2-\frac{r_n}{{\sqrt{q}}(1-q)}\right)
=\bt_{n+1}R_{n+1}R_n-\bt_nR_nR_{n-1},\nonumber
\eea
where the solution with the initial conditions $r_0=\bt_0=0$ is
\bea
r_n^2-\frac{r_n}{\sqq(1-q)}=\bt_nR_nR_{n-1}.
\eea
This expresses $\bt_n$ in terms of the subsidiary quantities $r_n$ and $R_n,$
\bea
\bt_n&=&\frac{r_n}{R_nR_{n-1}}\left(r_n-\frac{1}{\sqq(1-q)}\right)\non\\
&=&q^{-4n}-q^{-3n}.
\eea

In the next section we take a route for the computations of the
recurrence coefficients which does not involve the determination of the
analogous $r_n$ and $R_n.$

\setcounter{equation}{0}
\section{$q$-Laguerre polynomials}

This is also associated with an indeterminate moment problem at a level "higher"
than the Stieltjest-Wigert polynomials, in the sense that when an appropriate limit of a
parameter is taken, the $q-$Laguerre polynomials become the Stieltjest-Wigert polynomials.
See \cite{Koe:Swa}. We refer the reader to \cite{Mo} for further information.

 We take the weight to be,
\bea
w(x)=\frac{x^{\al}}{(-x;q)_\infty},\;\;0< x<\infty,\;\;\al>-1,\;\;0<q<1,
\eea
where
\bea
(z;q)_\infty:=\prod_{k=0}^{\infty}(1-zq^k)\non
\eea
This weight leads to
\bea
u(x)&=&\frac{q}{1-q}\left(\frac{1-q^{-\al}}{x}+\frac{q^{-\al}}{x+q}\right)\non\\
\frac{u(qx)-u(y)}{qx-y}&=&\frac{1}{1-q}\left(\frac{q^{-\al}-1}{xy}-
\frac{q^{-\al}}{(x+1)(y+q)}\right).\non
\eea
By definition,
\bea
A_n(x)&=&\frac{q^{-\al}-1}{\zeta_n(1-q)x}\int_{0}^{\infty}P_n(y)P_n(y/q)\frac{w(y)}{y}dy\non\\
&-&\frac{q^{-\al}}{\zeta_n(1-q)(x+1)}\int_{0}^{\infty}
P_n(y)P_{n}(y/q)\frac{w(y)}{y+q}dy\non\\
&=:&\frac{R_n}{x}-\frac{q^{n}}{(1-q)(x+1)}\\
B_n(x)&=&\frac{q^{-\al}-1}{(1-q)\zeta_{n-1}x}\int_{0}^{\infty}P_n(y)P_{n-1}(y/q)
\frac{w(y)}{y}dy\non\\
&-&\frac{q^{-\al}}{\zeta_{n-1}(1-q)(x+1)}\int_{0}^{\infty}
P_n(y)P_{n-1}(y/q)\frac{w(y)}{y+q}dy\non\\
&=:&\frac{r_n}{x}-\frac{q^{n-1}\textsf{p}_1(n)}{x+1},
\eea
where the evaluation of the second integrals in $A_n(x)$ and $B_n(x)$ will follow
later.

 Here is a computation of $R_0.$ By definition,
\bea
R_0&:=&\frac{q^{-\al}-1}{1-q}\;\frac{\int_{0}^{\infty}w(y)dy/y}{\int_{0}^{\infty}w(y)dy}\non\\
&=&\frac{q^{-\al}-1}{1-q}\;\frac{I(\al)}{I(\al+1)}=\frac{1}{1-q},\non
\eea
since
\bea
I(\al):=\int_{0}^{\infty}\frac{y^{\al-1}}{(-y;q)_\infty}dy=\frac{(q^{1-\al};q)_\infty}{(q;q)_\infty}
\frac{\pi}{\sin\pi\al}.\non
\eea
Also note the identity,
\bea
\frac{1}{(-x;q)_\infty(x+q)}=\frac{1}{q(1+x/q)(-x;q)_\infty}=\frac{1}{q(-x/q;q)_\infty}.\non
\eea
Hence,
\bea
\int_{0}^{\infty}P_n(y)P_n(y/q)\frac{w(y)}{y+q}dy
&=&\int_{0}^{\infty}P_n(y)P_n(y/q)\frac{y^{\al}}{q(-y/q;q)_\infty}dy\non\\
&=&q^{\al}\int_{0}^{\infty}P_n(qy)P_n(y)w(y)dy=q^{n+\al}\zeta_n,\non
\eea
and the result for $A_n(x)$ follows.
Similarly,
\bea
\int_{0}^{\infty}P_n(y)P_{n-1}(y/q)\frac{w(y)}{y+q}dy&=&\int_{0}^{\infty}P_n(y)P_{n-1}(y/q)
\frac{y^{\al}}{q(-y/q;q)_\infty}dy\non\\
&=&q^{\al}
\int_{0}^{\infty}P_n(qy)P_{n-1}(y)w(y)dy.\non
\eea
To complete the evaluation of the above integral, we note the
identity,
\bea
P_n(qy)&=&P_n(qy)+q^nP_n(y)-q^nP_n(y)\non\\
&=&q^nP_n(y)+q^ny^n+q^{n-1}\textsf{p}_1(n)y^{n-1}+...-q^n
\left(y^n+\textsf{p}_1(n)y^{n-1}+...\right)\non\\
&=&q^nP_n(y)+\textsf{p}_1(n)(q^{n-1}-q^{n})y^{n-1}+...\non\\
&=&q^nP_n(y)+\textsf{p}_1(n)(q^{n-1}-q^{n})P_{n-1}(y)+...\non
\eea
Finally,
\bea
&{}& q^{\al}\int_{0}^{\infty}P_n(qy)P_{n-1}(y)w(y)dy  \non\\
&=&q^{\al}\int_{0}^{\infty}
\{q^nP_n(y)+\textsf{p}_1(n)(q^{n-1}-q^n)P_{n-1}(y)+...\}P_{n-1}(y)w(y)dy\non\\
&=&\left(\frac{1}{q}-1\right)\textsf{p}_1(n)q^{n+\al}\zeta_{n-1},
\non
\eea
and the result for $B_n(x)$ follows.

 It turns out that for the $q-$Laguerre weight, the supplementary conditions
 produce 6 difference equations in contrast with the 4 in the previous example.

Now, by equating the residues
for the simple poles at $x=0$ and $x=-1$ in  \eqref{eqqS_1}, we find,
\bea
r_{n+1}+r_n&=&-\al_nR_n-\frac{1-q^{-\al}}{1-q}\\
\textsf{p}_1(n+1)q^{n}+\textsf{p}_1(n)q^{n-1}&=&-\frac{1+\al_n}{1-q}q^{n}
+\frac{1-q^{n+1}}{1-q}+\frac{q^{-\al}}{1-q},
\eea
respectively. We note here another identity involving $R_n$ only, by
equating the constant terms of \eqref{eqqS_1} at $x=\infty,$
\bea
R_n-\frac{q^{n}}{1-q}+(q-1)S_n-\frac{1-q^{n+1}}{1-q}=0,
\eea
where $S_n=\sum_{j=0}^{n}R_j.$
A similar consideration on  \eqref{eqqS_2} shows that
\bea
&{}& \al_n(r_n-r_{n+1})=\bt_{n+1}R_{n+1}-\bt_nR_{n-1}\\
&{}& \; -(1+\al_n)q^{n}{\sf p}_1(n+1)+(q+\al_n)q^{n-1} {\sf p}_1(n)
= \frac{\bt_{n+1}q^{n+1}-\bt_nq^{n-1}}{1-q}\\
&{}& r_{n+1}-qr_n = -q^{n}\al_n-1.
\eea

We use the fact that
$\al_n=\textsf{p}_1(n)-\textsf{p}_1(n+1)$ to rewrite (4.5) as a first order difference equation,
\bea
\textsf{p}_1(n+1)q^{n+1}-\textsf{p}_1(n)q^{n-1} = q^{n}+q^{n+1}-(1+q^{-\al})\non
\eea
which has an "integrating factor" $q^{n-1}$ by inspection. Hence the above equation becomes,
\bea
\quad \textsf{p}_1(n+1)q^{2n}-\textsf{p}_1(n)q^{2n-2}=(1+q)q^{2n-1}-(1+q^{-\al})q^{n-1},
\eea
and we find via a telescopic sum and the initial condition $\textsf{p}_1(0)=0,$
\bea
\textsf{p}_1(n+1)q^{2n}
&=&\frac{1-q^{2n+2}}{q(1-q)}-(1+q^{-\al})\frac{1-q^{n+1}}{q(1-q)}.\non
\eea
Therefore,
\bea
(1-q){\sf p}_1(n)&=&-q+(1+q^{-\al})q^{-n+1}-q^{-2n-\al+1},
\eea
and equation \eqref{eqp1intermsofalpha} gives
\bea
\qquad \al_n={\sf p}_1(n)-{\sf p}_1(n+1)=q^{-2n-1-\al}(1+q-q^{n+1}-q^{n+\al+1}).
\eea
At this stage $R_n$ can be found from a difference equation obtained by
subtracting (4.6) at "$n$" from the same at "$n+1$",
\bea
qR_{n+1}-R_n=q^{n+1}-q^{n},
\eea
with the initial condition $R_0=1/(1-q).$ Having now determined $\al_n$ in (4.12),
$r_n$ can be found from (4.9) with the initial condition $r_0=0.$ We proceed to
the determination of $\bt_n.$ Multiply (4.8) by the
"integrating factor" $q^n$ and by $1-q,$
\bea
&-&(1+\al_n)q^{2n}(1-q){\sf p}_1(n+1)+(q+\al_n)q^{2n-1}(1-q){\sf p}_1(n)\nonumber\\
&=&\bt_{n+1}q^{2n+1}-\bt_nq^{2n-1}.
\eea
The l.h.s. of the above is simplified to,
\bea
(1-q)q^{2n}\left({\sf p}_1(n)-{\sf p}_1(n+1)\right)
-\al_n(1-q)\left({\sf p}_1(n+1)q^{2n}-{\sf p}_1(n)q^{2n-1}\right)\nonumber\\
=(1-q)q^{2n}\al_n\left[1-\left({\sf p}_1(n+1)-{\sf p}_1(n)/q\right)\right]\nonumber.
\eea
With (4.11), the term ${\sf p}_1(n+1)-{\sf p}_1(n)/q$ simplifies to
\bea
1-q^{-2n-\al-1},\nonumber
\eea
and consequently (4.14) reduces to the first order difference equation, with the initial condition
$\bt_0=0,$
\bea
\bt_{n+1}q^{2n+1}-\bt_nq^{2n-1}=(1-q)q^{-1-\al}\al_n.
\eea
Taking a telescopic sum, and noting that
$
\sum_{j=0}^{n-1}\al_j=-\textsf{p}_1(n),
$
\bea
\bt_nq^{2n-1}&=&(1-q)q^{-1-\al}\sum_{j=0}^{n-1}\al_n
=-(1-q)q^{-1-\al}{\sf p}_1(n).
\eea
Finally,
\bea
\bt_n&=&-q^{-2n-\al}(1-q)\textsf{p}_1(n)\non\\
&=&-q^{-2n-\al}(-q+(1+q^{-\al})q^{1-n}-q^{-2n-\al+1})\nonumber\\
&=&q^{-4n-2\al+1}(1-q^n)(1-q^{n+\al}).
\eea
It is interesting to note that in the computations of $\al_n$ and $\bt_n,$ not
all the 6 equations are required. We have used only (4.5) and (4.8).
However, for an explicit expression of the $q-$Ladder operators and therefore
the determination of $r_n$ and $R_n$ we need the equations (4.6), (4.9) and $\al_n$ in (4.12).

We end with the remark that in the case of the classical Laguerre polynomials,
$\al_n=2n+1+\al,$ and $\bt_n=\sum_{j=0}^{n-1}\al_j=n(n+\al),$ and this is analogous
to (4.12) and (4.16)-(4.17), however, with appropriate modifications in the $q-$case.


\begin{thebibliography}{99}

\bibitem{Akh}N. I. Akhiezer, {\em The Classical Moment Problem and Some Related
Questions in Analysis}, English translation, Oliver and Boyed, Edinburgh,
1965.
%
\bibitem{Bau} W. Bauldry, Estimates of asymptotic
Freud polynomials on the real line, {\it J. Approx. Theory}
{\bf 63} (1990) 225--237.
%
\bibitem{Bon:Cla} S. S. Bonan and D. S. Clark,
Estimates of the Hermite and Freud polynomials, {\it J.
Approx. Theory} {\bf 63} (1990), 210--224.
%
\bibitem{Che:Ism} Y. Chen and M. E. H. Ismail, {\it Ladder
operators and differential equations for orthogonal
polynomials,} J. Phys. A {\bf 30} (1997), 7818--7829.
%
\bibitem{Che:Ism2} Y. Chen and M. E. H. Ismail,
{\it Jacobi polynomials from compatibility conditions,}
  Proc. Amer. Math. Soc. {\bf 133} (2004), 465-472.
%
\bibitem{Chr03} J. S. Christiansen, {\it The moment problem associated
with the Stieltjes--Wigert polynomials,} J. Math. Anal. Appl.
{\bf 277} (2003), 218--245.
%
\bibitem{Ism8}, M. E. H. Ismail, {\it Difference equations and quantized
discriminants for $q$-orthogonal polynomials,}
Advances in Applied Math. {\bf 30} (2003), 562--589.
%
\bibitem{Ismbook} M. E. H. Ismail, {\em Classical and Quantum Orthogonal
Polynommials in one Variable}, Cambridge University Press,
Cambridge, 2005.
%
\bibitem{Ism:Nik:Sim} M. E. H. Ismail, I. Nikolova and P. Simeonov,
{\it Difference equations and discriminants for discrete orthogonal polynomials,
the Ramanujan Journal,} {\bf 8} (2004), 475--502.
%
\bibitem{Koe:Swa} R. Koekoek and R. Swarttouw, {\em The Askey-scheme of
hypergeometric orthogonal polynomials and its $q$-analogues,
Reports of the Faculty of Technical Mathematics and Informatics
no. 98-17, Delft University of Technology, Delft,} 1998.
%
\bibitem{Mha} H. L. Mhaskar, {\it Bounds for certain Freud polynomials,}
J. Approx. Theory {\bf 63} (1990),  238--254.
%
\bibitem{Mo} D. S. Moak, The $q-$analogue of the Laguerre polynomials,
{\bf 81} (1981), 20--47.
%
\bibitem{Sze} G. Szeg\H{o}, {\it Orthogonal
Polynomials}, 4th edition, Amer. Math. Soc., Providence, 1975.
%



\end{thebibliography}
\end{document}